\documentclass[runningheads]{svmult}

\usepackage{makeidx}   
\usepackage{graphicx}  
\usepackage{subeqnar}  
\usepackage{multicol}  
\usepackage{physprbb}  
\makeindex             

\begin{document}
\title*{RR Lyrae Distance Scale: Theory and Observations}
\toctitle{RR Lyrae Distance Scale}
\titlerunning{RR Lyrae Distance Scale}
\author{Giuseppe Bono}
\authorrunning{Giuseppe Bono}
\institute{INAF - Rome Astronomical Observatory, 
           Via Frascati 33,\\
           00040 Monte Porzio Catone, Italy}    

\maketitle              

\begin{abstract}

The RR Lyrae distance scale is reviewed. In particular, we discuss 
theoretical and empirical methods currently adopted in the literature. 
Moreover, we also outline pros and cons of optical and near-infrared 
mean magnitudes to overcome some of the problems currently affecting 
RR Lyrae distances. The importance of the K-band 
Period-Luminosity-Metallicity ($PLZ_K$) relation for RR Lyrae is also 
discussed, together with the absolute calibration of the zero-point. 
We also mention some preliminary results based on NIR (J,K) time 
series data of the LMC cluster Reticulum. This cluster hosts a sizable 
sample of RR Lyrae and its distance is found to be $18.45\pm 0.04$ mag 
using the predicted $PLZ_K$ relation and $18.51\pm 0.06$ using the 
$PLZ_J$ relation. We briefly discuss the evolutionary status of 
Anomalous Cepheids and their possible use as distance indicators. 
Finally, we point out some possible improvements to improve the 
intrinsic accuracy of theory and observations.   
\end{abstract}

\section{Introduction}

During the last half a century RR Lyrae stars have been the crossroad 
of paramount theoretical and observational efforts. The reasons are 
manifold. From a theoretical point of view RR Lyrae play a crucial 
role because they are a fundamental laboratory not only to test the 
accuracy of evolutionary (Cassisi et al.\cite{cas99}  Brown et al.\cite{br01}; 
VandenBerg \& Bell\cite{vdb01}) and pulsation (Bono \& Stellingwerf\cite{bs94};
Bono et al.\cite{bccm97}; Feuchtinger\cite{f99}) models but also to constrain 
fundamental physics problems such as the neutrino magnetic moment 
(Castellani \& Degl'Innocenti\cite{cdi93}).

From an observational point of view RR Lyrae are even more important 
since they the are most popular primary distance indicators for old, 
low-mass stars (see e.g. Smith\cite{s95}; Caputo\cite{c98}; 
Walker\cite{w99}\cite{w00}; Carretta et al.\cite{car00}; Walker, these 
proceedings; Cacciari \& Clementini these proceedings). Dating back to 
Baade\cite{b44}\cite{b58}, RR Lyrae stars have been also adopted to 
trace the old stellar component in the Galaxy (Suntzeff et al.\cite{s91}; 
Layden\cite{l95}) and in nearby galaxies (Mateo\cite{m98}; 
Monelli et al.\cite{mmo03}). The use of RR Lyrae as stellar tracers 
received during the last few years a new spin. On the basis of time-series 
data recent photometric surveys identified a local overdensity of RR Lyrae 
stars in the Galactic halo (Vivas et al.\cite{v01}). Current empirical 
(Yanny et al.\cite{y00}; Ibata et al.\cite{i01}; Martinez-Delgado et 
al.\cite{m01}) and theoretical (Helmi\cite{h02}, and references therein) 
evidence suggests that such a clump is the northern tidal stream left over 
by the Sagittarius dwarf spheroidal (dSph). 

This not withstanding, several empirical phenomena connected with the 
evolutionary and pulsation properties of RR Lyrae stars have not been 
settled yet. We still do not know the physical mechanisms that govern 
the occurrence of the {\em Blazhko effect} (Kolenberg et al.\cite{k03};
Smith et al.\cite{s03}) as well as of the mixed-mode behavior 
(Bono et al.\cite{bccm96}; Feuchtinger\cite{f98}). The same outcome 
applies for the formation and propagation of the shock front along the 
pulsation cycle (Bono et al.\cite{bcs94b}; Chadid et al.\cite{ch00}).  

However, the lack of a detailed knowledge of the physical phenomena 
that take place in the interior, in the envelope, and in the atmosphere 
of RR Lyrae stars only partially hampers the use of these objects as 
standard candles. In the following we discuss pros and cons in using 
different theoretical and empirical relations to derive RR Lyrae 
distances. In particular, we will focus our attention on optical 
and near-infrared (NIR) data for field and cluster RR Lyrae. Finally, 
we briefly outline the current status of RR Lyrae and classical Cepheid 
distance scales.

\section{Theoretical and Empirical Circumstantial Evidence}

At present, the most popular approach to estimate the RR Lyrae distances 
is the $M_V$-[Fe/H] relation. This relation is widely adopted because 
it only requires two observables, namely the apparent visual magnitude 
and the metallicity. From a theoretical point of view it is also 
well-defined, because the RR Lyrae instability strip is located in 
a region of the Horizontal Branch (HB) that is quite flat. In spite of 
these straightforward positive features the absolute calibration of the 
$M_V$-[Fe/H] relation is still an open problem. Current theoretical and 
empirical calibrations provide difference in absolute distances that 
range from 0.1 to 0.25 mag (Bono et al.\cite{bccm01}). Oddly enough, 
the internal 
errors are quite often of the order of a few hundredths of magnitude.   
This indicates that current methods might be affected by deceptive 
systematic errors. The main problems affecting the $M_V$-[Fe/H] relation 
are the following:

{\em i) Evolutionary effects.} The use of the $M_V$-[Fe/H] relation 
relies on the assumption that RR Lyrae stars are on the 
Zero-Age-Horizontal-Branch (ZAHB). This is on average a plausible but thorny 
assumption, since field and cluster RR Lyrae do show a spread in luminosity. 
Moreover, theoretical (Bono et al.\cite{bcd95}; Cassisi \& Salaris\cite{cas97})
 and empirical 
(Carney, Storm, \& Jones\cite{ca92}; Sandage\cite{sa93}) evidence 
suggest that the intrinsic width in luminosity of the ZAHB becomes larger 
when moving from metal-poor to metal-rich Galactic Globular Clusters (GGCs). 
As a consequence, RR Lyrae samples at different metal contents are 
differentially affected by off-ZAHB evolution as well as by the HB morphology 
(Caputo\cite{c98}, and references therein). Moreover, a spread in luminosity of 
the order of $\pm0.1$ dex causes a spread in the visual magnitude of the 
order of $\pm0.25$ mag (see Fig. 1 in Bono et al.\cite{bccm01}). To 
investigate in more detail this effect we estimated, using the atmosphere 
models provided by Castelli, Gratton, \& 
Kurucz\cite{cgk97a}\cite{cgk97b}\footnote{The models 
labeled {\em NOVER} were constructed by adopting a canonical treatment for 
the mixing-length, i.e. the convective overshooting into the convective 
stable regions is neglected.}, the bolometric 
correction in the V-band for two different ZAHBs that cover the metallicity 
range typical of Galactic RR Lyrae. According to current evolutionary 
predictions we adopted $\log L/L_\odot=1.75$, $M/M_\odot=0.75$ for Z=0.0001 
and  $\log L/L_\odot=1.51$, $M/M_\odot=0.55$ for Z=0.02. Data plotted in the 
top panel of Fig. 1 show that spread in luminosity is mainly due to the 
change in the bolometric correction. When moving from the hot (blue) to the 
cool (red) edge of the instability strip $BC_V$ undergoes a changes of 
the order of 0.1 mag\footnote{Note that we assumed a range in temperature 
of more than 3,000 K to account for the temperature variation along the 
pulsation cycle.}. Therefore, RR Lyrae stars with exactly the same stellar 
mass and luminosity but different effective temperatures (pulsation periods) 
present an intrinsic spread in the visual magnitude of approximately a tenth 
of a magnitude. Data plotted in the top panel are also suggesting that 
the ZAHB luminosity, at fixed bolometric magnitude, tilts when moving 
toward hotter effective temperatures (Brocato et al.\cite{broc99}).  

This trend presents a substantial change when moving toward longer wavelengths, 
and indeed the bolometric correction in the I-band provides for the same ZAHBs 
a change of the order of 0.3-0.4 mag. This means 
that GGCs characterized by well-populated instability strips start to display 
a slope when moving from hotter to cooler objects. Therefore, cluster RR Lyrae 
stars with the same stellar mass and luminosity become systematically brighter 
when moving from shorter to longer periods. It turns out that RR Lyrae in the 
I-band start to obey a Period-Luminosity (PL) relation. This effect 
become more and more evident once we move from the I to the K-band, and 
indeed the bolometric correction increases by more than one magnitude when 
moving from the blue to the red edge of the instability strip. Therefore, 
RR Lyrae stars should show a well-defined PL relation in the K-band. 
This result strongly supports the seminal empirical finding brought forward 
by Longmore et al.\cite{l90} concerning the occurrence of the K-band 
PL relation among cluster RR Lyrae. Moreover, the $PL_K$ should also 
be marginally 
affected by the intrinsic spread in luminosity, since a mild decrease 
in the effective temperature (increase in the period) causes a strict 
increase in brightness, and in turn a decrease in $M_K$.  
It is worth mentioning that in performing this test we adopted the same 
evolutionary predictions (bolometric magnitudes and effective temperatures) 
and that the change from the V to the K-band is mainly due to the 
bolometric corrections and the color-temperature relations predicted by 
atmosphere models. Finally we note that RR Lyrae $M_K$ magnitudes, in 
contrast with $M_V$ magnitudes, are also marginally affected by a spread 
in stellar mass inside the instability strip (see Fig. 2 in Bono et al. 
\cite{bccm01}).

\begin{figure}[ht]
\begin{center}
\includegraphics[width=.8\textwidth]{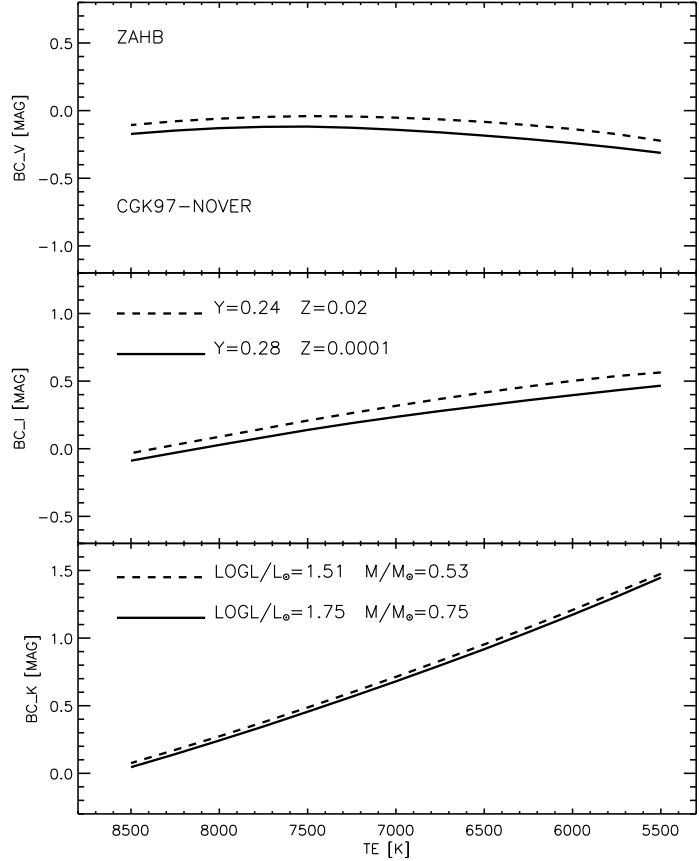}
\end{center} 

\vspace*{0.75truecm} 
\caption[]{Top: predicted bolometric correction in the visual band 
as a function of the effective temperature. The solid line show the 
change of $BC_V$ along the ZAHB for very metal-poor RR Lyrae stars 
(Z=0.0001), while the dashed line is for metal-rich ones (Z=0.02). 
The $BC_V$ values were estimated using the atmosphere models with no 
overshooting (NOVER) provided by CGK97. Middle: same as the top, but
the bolometric corrections refer to the I-band. Bottom: same as 
the top, but the bolometric corrections refer to the K-band.}
\label{Fig1}
\end{figure}

{\em ii) Linearity.} Recent theoretical and empirical evidence indicates   
that the $M_V$-[Fe/H] relation is not linear when moving from metal-poor 
to metal-rich RR Lyrae (Castellani, Chieffi, Pulone\cite{ccp91}; 
Caputo et al.\cite{ccm00}; Layden\cite{la02}). In particular, the slope 
appears to be quite shallow in the metal-poor regime 
(0.18 for $[Fe/H]\le-1.6$) while it is quite steep in the metal-rich 
regime (0.35 for $[Fe/H]>-1.6$). However, recent photometric and spectroscopic 
measurements of RR Lyrae stars in the Large Magellanic Cloud (LMC) do not 
show evidence of a change in the slope at $[Fe/H]\approx-1.5$ 
(Clementini et al.\cite{cle03}). The change in the slope, if confirmed 
by new and independent estimates, means that metal abundance might also 
introduce a systematic uncertainty when moving from metal-poor to 
metal-rich RR Lyrae. In fact, an uncertainty of $\pm0.2$ dex in 
metallicity implies an uncertainty in visual magnitude that ranges from 
0.04 to 0.07 mag. Note that such an uncertainty would affect not only 
the slope but also the zero-point of the $M_V$-[Fe/H] relation.  
On the other hand, data plotted in the bottom panel of Fig. 1 and in 
Fig. 3 by Bono et al.\cite{bccm01}, together with K-band observational 
data (Longmore et al.\cite{l90}) suggest that the $PL_K$ presents a 
linear dependence on the metal content. Note that  

{\em iii) Reddening.} It is well-known that an uncertainty in the 
reddening, E(B-V), of 0.01 mag 
implies an uncertainty of the  order of 0.03 mag in the visual magnitude.
The same error in the reddening causes an uncertainty that is a factor
of two smaller in the I-band, and negligible in the K-band (Cardelli et 
al.\cite{ccm89}). Moreover, new empirical optical (B-V, V-I) relations 
based on cluster (Kovacs \& Walker\cite{kw01}) and field 
(Piersimoni, Bono, \& Ripepi\cite{pbr02}) 
fundamental mode RR Lyrae stars might supply accurate individual reddening 
estimates. Interestingly enough, these relations rely on reddening 
free parameters, such as period, luminosity amplitude, and metallicity 
and present a small intrinsic dispersion.  

{\em iv) Mean Magnitudes.} The mean magnitude of RR Lyrae stars is 
estimated as time average either in magnitude or in intensity along 
the pulsation cycle. However, current theoretical (Bono, Caputo, \& 
Stellingwerf\cite{bcs94a}; Marconi et al.\cite{mdc03}) and 
empirical (Corwin \& Carney\cite{cc01}) evidence suggest that the 
two mean magnitudes present a systematic difference with the mean 
"static" magnitude of equivalent nonpulsating stars. The discrepancy for 
fundamental RR Lyrae stars ($RRab$) increases from a few hundredths of 
a magnitude close to the red edge to $\approx 0.1$ mag close to the blue 
edge. This discrepancy becomes marginal in the K-band, since the luminosity
amplitude becomes a factor of $\approx 3$ smaller than in the V-band.   
    
{\em v) Metallicity.} Recent spectroscopic investigations based 
on high-resolution spectra collected with 8m-class telescopes disclosed 
that hot HB stars present a quite complicate pattern of both helium and 
heavy element abundances (Behr et al. \cite{be99}; Moehler et al. 
\cite{mo02}). These peculiarities have also been identified as jumps 
along the ZAHB in the near ultraviolet bands (Grundhal et al. \cite{gr99}; 
Momanhy et al.\cite{mom02}). According to current beliefs these peculiar 
abundances are the balance between two competing effects, namely gravitational  
settling and radiative levitation (Michaud, Vauclair, \& 
Vauclair\cite{mvv83}). Up to now only for a few field RR Lyrae are 
available high resolution spectra (Clementini et al.\cite{cle95}), and  
therefore we do not know whether cluster RR Lyrae present the same 
chemical peculiarities. Moreover, the metallicity of cluster RR Lyrae 
is generally estimated using the $\Delta S$ method or the {\em hk} 
index (Anthony-Twarog et al.\cite{at91}; Rey et al.\cite{re00}) but 
both of them are based on the Ca abundance that is an $\alpha-element$.   
The empirical scenario was further jazzed up by the evidence that the 
stellar rotation shows a bimodal distribution in the hot region of 
the HB (Recio-Blanco et al.\cite{rec02}). 
On the other hand, current empirical evidence indicate that RR Lyrae 
stars do not rotate rapidly enough for the rotation to be detected 
(Peterson, Carney, \& Latham\cite{pcl96}).      

Together with this is the indisputable fact we are still facing the 
problem of the metallicity scale. In fact, the Zinn \& West\cite{zw84} 
and the Carretta \& Gratton\cite{cg97} scales show in the intermediate 
metallicity range a difference of the order of 0.2 dex (Rutledge, 
Hesser, \& Stetson\cite{rhs97}; Kraft \& Ivans\cite{ki03}). Note 
that the zero-point of the 
$M_V$-[Fe/H] relation is generally estimated at [Fe/H]=-1.5, and 
therefore current uncertainties on the metallicity scale might 
introduce an error of the order of 0.04 mag(!).  
Finally, we mention that we still lack a detailed knowledge of 
$\alpha-element$ abundances among cluster RR Lyrae stars. This 
parameter is crucial to estimate the global metallicity, i.e. 
the metallicity currently adopted in constructing both evolutionary 
and pulsation models (Salaris, Chieffi, \& Straniero\cite{scs93}; 
Zoccali et al.\cite{z00}). According 
to current empirical evidence the overabundance of $\alpha-elements$  
should decrease when moving from metal-poor to metal-rich stars 
(Carney\cite{ca96}), but current spectroscopic data for RR Lyrae 
stars are scanty.

Obviously the abundance of $\alpha-elements$ also affects atmosphere 
models, and in turn the BCs and the color-temperature (CT) relations. 
We performed a test using the $\alpha-enhanced$ atmosphere models  
recently constructed by Castelli et al. (2003\footnote{These models 
as well as the Castelli et al.\cite{cgk97a} models are available at 
the following web site http://kurucz.harvard.edu}) assuming an 
$\alpha$ over iron enhancement of $[\alpha/Fe]\approx0.4$.   
We found that at fixed iron abundance ($-2.0 \le [Fe/H]\le -1$) the 
difference in BCs and in CTs,  for surface gravities ($\log g=2.5-3.0$) 
and effective temperatures ($5300\le T_e \le 8300$ K) typical of 
RR Lyrae stars, between solar-scaled and $\alpha-enhanced$ models is 
small and of the order of a few hundredths of magnitude.

\begin{figure}[ht]
\begin{center}
\includegraphics[width=.8\textwidth]{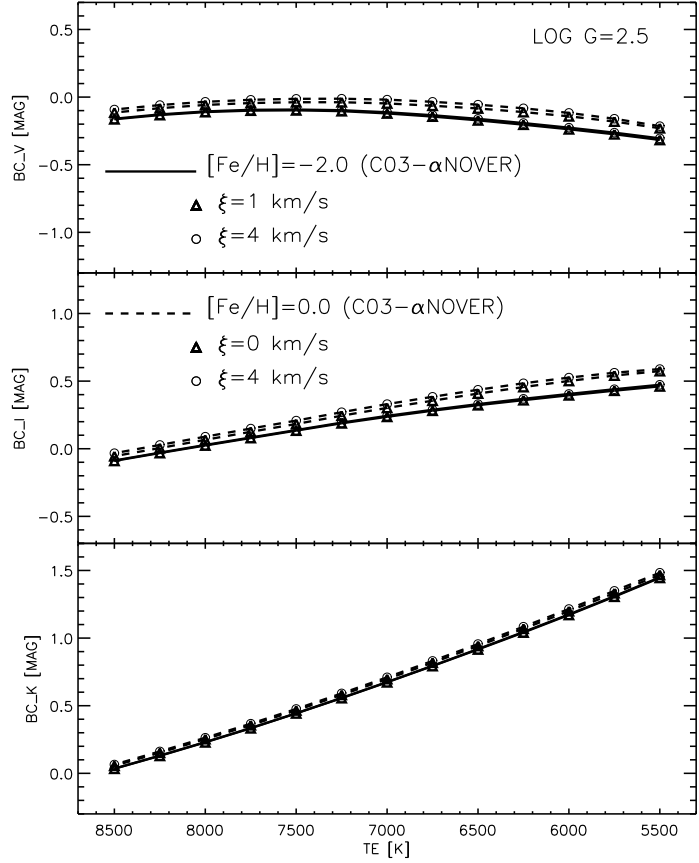}
\end{center} 

\vspace*{0.75truecm} 
\caption[]{Top: predicted bolometric correction in the visual band 
as a function of the effective temperature. The solid and the dashed 
lines display the change of $BC_V$ at fixed gravity ($\log g =2.5$) 
for metal-poor ([Fe/H]=-2.0) and metal-rich ([Fe/H]=0.0) RR Lyrae 
stars. The atmosphere models (C03) adopted to estimate the $BC_V$ 
values were constructed by adopting an overabundance of $\alpha$-elements 
of $[\alpha/Fe]=0.4$ and different assumptions for the microturbulent 
velocity (see labeled values). Middle: same as the top, but the bolometric 
corrections refer to the I-band. Bottom: same as the top, but the bolometric 
corrections refer to the K-band.}
\label{Fig2}
\end{figure}

{\em vi) Microturbulent velocity.} Theoretical and empirical evidence 
suggests that the microturbulent velocity ($\xi$) in the atmosphere of RR Lyrae 
stars ranges from a few km/s to more than 10 km/s along the pulsation 
cycle and peaks close to the phases of maximum compression (Benz \& 
Stellingwerf\cite{bes85}; Cacciari et al.\cite{ccpb89}; Fokin, Gillet, 
\& Chadid\cite{fgc99}).  
The sample of RR Lyrae for which this information are available is quite 
limited; however, current data seem to indicate that the microturbulent 
velocity is larger than 5 km/s for a substantial fraction of the pulsation 
period. 

On the other hand, current atmosphere models are constructed by adopting 
a microturbulent velocity of 2 km/s, since this is a typical value for 
static stars. As a consequence, evolutionary and pulsation predictions 
when transformed into the observational plane might be affected by a 
systematic uncertainty. Therefore we decided to investigate the 
dependence of both BCs and CTs on this fundamental parameter. 
Fig. 2 shows the variation of BCs, at fixed surface gravity ($\log g=2.5$),
for two sets of $\alpha-enhanced$ atmosphere models constructed by 
adopting different iron abundances and microturbulent velocities
(see labeled values)\footnote{Current models are also available in the 
Kurucz web site. Note that for [Fe/H]=-2.0 we adopted $\xi=1$ and 4 km/s, 
because $\alpha-enhanced$ atmosphere models for $\xi=0$ are not available 
yet.}. Data plotted in the top panel show quite clearly that $BC_V$ 
values of metal-poor models are marginally affected by this parameter 
inside the instability strip, while the metal-rich ones present a 
difference of the order of a few hundredths of magnitude. The same outcome
applies for the $BC_I$s. Once again the BC in the K band shows marginal 
changes both in the metal-poor and in the metal-rich regime.

\begin{figure}[ht]
\begin{center}
\includegraphics[width=.8\textwidth]{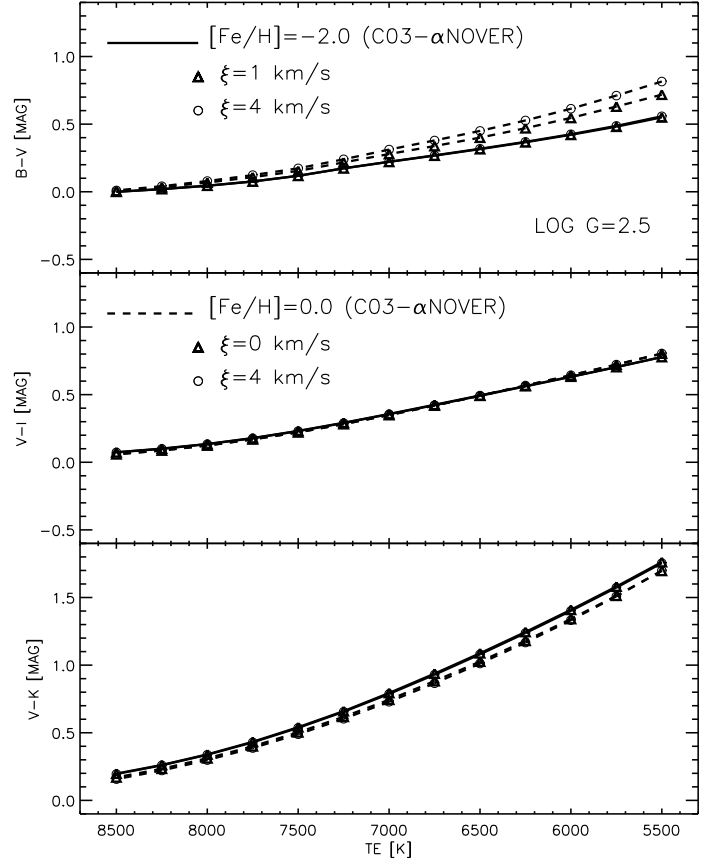}
\end{center} 

\vspace*{0.75truecm} 
\caption[]{Top: same as Fig. 1, but for the B-V color. Middle: same as the 
top, but for the V-I color. Bottom: same as the top, but for the V-K color.}
\label{Fig3}
\end{figure}

Let us now investigate the dependence of the CT relations on the 
microturbulent velocity. The top panel of Fig. 3 shows the B-V color 
as a function of the effective temperature for the same grid of atmosphere 
models adopted in Fig. 2. Metal-poor B-V colors present a marginal 
dependence on $\xi$ inside the instability strip. On the other hand, 
the models at solar chemical composition show that the difference between 
the models with $\xi=0$ and $\xi=4$ km/s strictly increases when moving 
from hotter to cooler effective temperatures. The difference, close 
to the red edge of the instability strip, becomes of the order of a tenth 
of magnitude (!). Interestingly enough, the V-I colors (middle panel) do 
not show any dependence at all on the metal abundance as well as on the 
microturbulent velocity, thus suggesting that in the V and the I-band the 
two effects cancel out. Finally, the V-K colors (bottom panel) show a 
mild {\em reversed} (Bono et al.\cite{bccm01}) dependence on metal abundance 
and a marginal dependence on the microturbulent velocity. In this 
context it is worth mentioning that Cacciari et al.\cite{cacl00} have recently 
revised the zero-point of the Baade-Wesselink (BW) method using a new set 
of atmosphere models that partially overlaps with the atmosphere models 
currently adopted. Using the entire set of photometric and spectroscopic 
data available in the literature for two fields RR Lyrae they found that 
the absolute magnitude of RR Cet ([Fe/H]=$-1.45$) is $\approx$ 0.12 mag 
brighter than previously estimated. However, they did not find any 
significant change between old and new estimates for SW And 
([Fe/H]=$-0.24$).  

In this section we discussed some possible uncertainties affecting  
the RR Lyrae distance scale. It is worth mentioning that several 
of them might affect both theory and observations, therefore it is 
quite difficult to estimate the global error budget on a quantitative 
basis. However, it turns out that sets of atmosphere models, constructed 
by adopting different physical assumptions, predict BCs that might differ 
by a few hundredths of magnitude. The impact on the B-V colors is larger 
and of the order of 0.1 mag.

\section{New Theoretical Approach}

The circumstantial evidence discussed in the previous section suggested 
that the  K-band PL relation of RR Lyrae should present several advantages 
when compared with the other methods currently adopted in the literature.  
Moreover, and even more importantly, Longmore et al.\cite{l90} demonstrated, 
on the basis of K-band photometry for a good sample of GGCs, that cluster 
RR Lyrae do obey to a well-defined PL relation.   
Therefore, we decided to investigate whether nonlinear, time-dependent 
convective models of RR Lyrae (Bono \& Stellingwerf\cite{bs94}) support 
this empirical scenario. 
To cover the metal abundances typical of Galactic RR Lyrae we computed 
several sequences of models ranging from Y=0.24, Z=0.0001 to Y=0.28, Z=0.02.  
For each given chemical composition we adopted a single mass-value and 
2-3 different luminosity levels to account for off-ZAHB evolution and for 
possible uncertainties on the ZAHB luminosity predicted by current 
evolutionary models. The main advantage in adopting this approach is that 
the edges of the instability region can be consistently estimated. 
Even though current predictions depend on the adopted mixing-lenght 
parameter, they do not rely on {\em ad hoc} assumptions concerning the 
position of the red edge (Bono et al.\cite{bms99}). 

We found that RR Lyrae models do obey to a well-defined $PLZ_K$ relation: 
\begin{equation}
M_K = 0.139 -2.071(\log P +0.30) + 0.167\log Z 
\end{equation} 
 
with an intrinsic dispersion of 0.037 mag. The symbols have their usual 
meaning. On the basis of this relation and of K-band data for RR Lyrae 
stars in M3 collected by Longmore et al.\cite{l90} we found for this 
cluster a true distance modulus of $15.07\pm0.07$ mag. This estimate 
is in very good agreement with the distance provided by Longmore et al., i.e.  
$DM=15.00\pm0.04\pm0.15$ mag, where the former error refers to uncertainties 
in the zero-point, while the latter in the slope of the PL relation.  
It is noteworthy that the quoted distances are also in good agreement 
with the M3 distance based on the First Overtone Blue Edge (FOBE) method  
developed by Caputo et al.\cite{ccm00}. This method is based on the comparison 
between the predicted first overtone blue edge and the location of $RRc$ 
variables in the $\log P$ vs $M_V$ plane. The accuracy of this method 
depends on the number of $RRc$ variables present in a given stellar 
system and seems to provide accurate distances for GCs characterized 
by well-populated instability strips. In the case of M3 they found 
$DM=15.00\pm0.07$ mag. 

Although the $PLZ_K$ relation for RR Lyrae presents several indisputable 
advantages, when compared with other methods available in the literature,   
we still lack accurate measurements of mean K-band magnitude for 
cluster RR Lyrae. Therefore, the comparison with empirical PL relations 
did not allow us to constrain the intrinsic accuracy of our predictions.  
Fortunately enough, Benedict et al.\cite{ben02} provided an accurate estimate 
of the trigonometric parallax of RR Lyr itself using FGS3, the interferometer 
on board of the Hubble Space Telescope (HST). Note that the new estimate,  
$\pi_{abs}=3.82\pm=0.20$ mas, is approximately a factor of three more 
accurate than the previous evaluation provided by {\em Hipparcos}, i.e.
$\pi_{abs}=4.38\pm=0.59$ mas. Therefore, we investigated whether the 
theoretical framework we developed accounts for this accurate absolute 
distance. By adopting for RR Lyr a mean interstellar extinction of 
$<A_V>=0.12\pm0.1$ (Benedict et al.\cite{ben02}), an iron 
abundance of [Fe/H]=-1.39 
($Z\approx0.0008$, Fernley et al.\cite{fer98}; Clementini et al.\cite{cle95}),  
a mean K magnitude $K=6.54\pm0.04$ mag (Fernley, Skillen, \& 
Burki\cite{fer93}), and a period of $\log P=-0.2466$ (Hardie\cite{h55}) 
we found a pulsation parallax of 
$\pi_{abs}=3.858\pm=0.131$ mas. The absolute distance we obtained 
agrees quite well with the new parallax for RR Lyr provided by HST.   
This result, once confirmed by new and accurate geometrical distances, 
emphasizes the potential of the $PLZ_K$ in view of a new NIR RR Lyrae 
distance scale.

\section{New Observational Approach}

We already mentioned that accurate mean K-band magnitudes are only 
available for a limited sample of cluster RR Lyrae (Liu \& Janes\cite{lj90}; 
Longmore et al.\cite{l90}; Storm et al.\cite{scl94}\cite{snc94}). 
These data are not very 
accurate, since NIR photometry with small format detectors was partially 
hampered by crowding. A few K-band measurements have also been collected 
by Carney et al.\cite{cf95} for RR Lyrae stars in the Galactic bulge.  
However, field RR Lyrae whose distances were estimated using the BW 
method (26 $RRab$ plus 3 RR Lyrae pulsating in the first 
overtones, $RRc$) have mean K-magnitudes 
with an accuracy of the order of a few hundredths of magnitude. 
A preliminary comparison between distances based on the BW method 
and on the $PLZ_K$ relation discloses a systematic difference 
that decreases when moving from metal-poor to metal-rich objects
(Bono et al.\cite{bon03}). It is worth mentioning that this discrepancy 
between the two different methods is substantially reduced once we 
adopt the new calibration of the BW method provided by 
Cacciari et al.\cite{cacl00}.    

It goes without saying that new and accurate mean K-band magnitudes 
for cluster RR Lyrae are mandatory to improve current theoretical 
and empirical scenarios. Therefore, we decided to start a new observational 
project aimed at collecting J and K band data in a dozen Galactic and 
Magellanic Cloud clusters. 
Figures 4 and 5 show the K,V-K and the J,V-J Color-Magnitude Diagram of 
the LMC cluster Reticulum. 
We selected this cluster because it contains a sizable sample of RR Lyrae 
(32) and it is characterized by a very low central density. Moreover, 
accurate periods for the entire sample are available in the literature
(Walker\cite{w92}).  

\begin{figure}[ht]
\begin{center}
\includegraphics[width=1.0\textwidth]{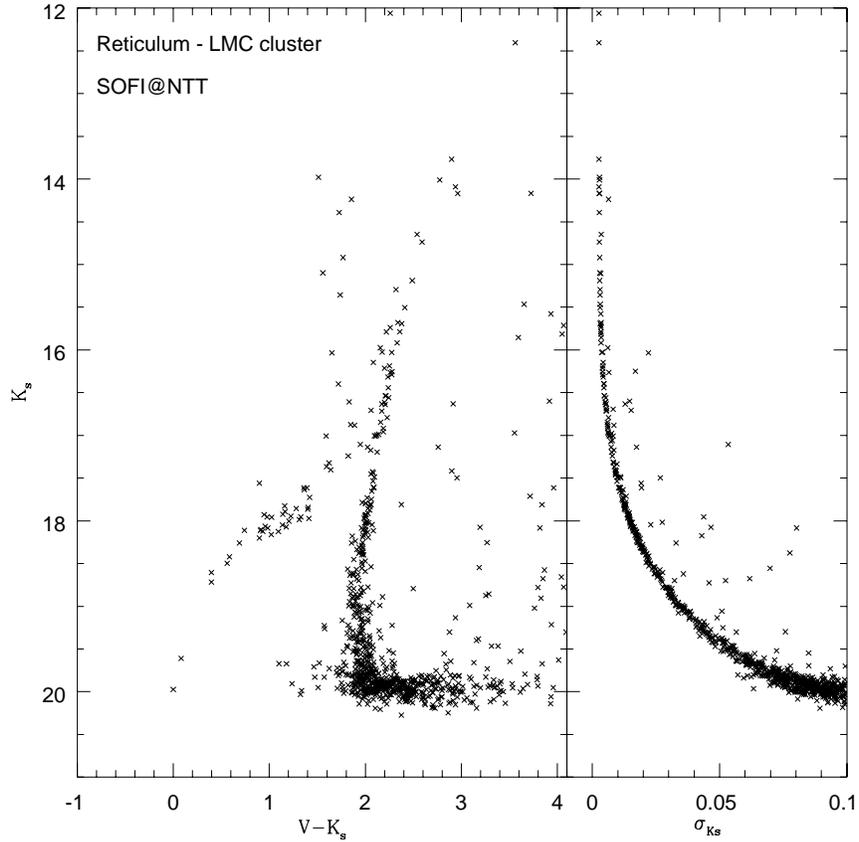}
\end{center} 
\caption[]{Left: Color magnitude diagram of the LMC cluster Reticulum 
in K,V-K. Data were collected over three different observing runs with
SOFI@NTT and reduced using DAOPHOT/ALLFRAME. A glance at the data shows 
that RR Lyrae stars in this cluster present a well-defined slope. Right: 
intrinsic photometric error. The strategy adopted to perform the photometry
allowed us to reach a K-band limiting magnitude of 19.5 with an accuracy 
better than 0.05 mag.}
\label{Fig4}
\end{figure}

Optical (UBVI) data were collected using SUSI2 at ESO/NTT, the NIR ones 
with SOFI at ESO/NTT and cover a time interval of three years. In summary, 
we collected approximately 170 phase points in the K-band and at roughly 50 
phase points in the J-band. The individual exposure times range from 
1 to 2 minutes in the K-band and from 20 s to 1 minute in the J-band. 
To improve the accuracy of individual measurements we adopted a new 
reduction strategy, i.e. we performed with DAOPHOT/ALLFRAME the 
photometry over the entire set of J and K individual exposures.

\begin{figure}[ht]
\begin{center}
\includegraphics[width=1.0\textwidth]{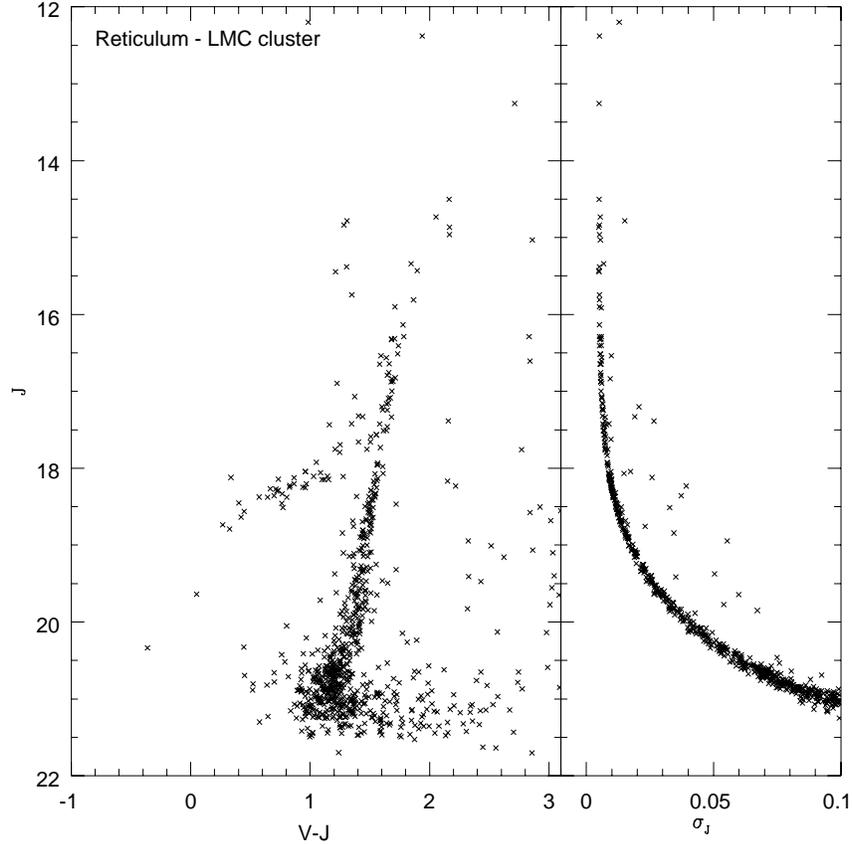}
\end{center} 
\caption[]{Left: same as Fig. 4, but the CMD is J,V-J. Right: intrinsic 
photometric error. Note that RR Lyrae stars also show a slope in the 
J-band but flatter than in the K-band. The strategy adopted to perform 
the photometry allowed us to reach a J-band limiting magnitude of 20.5 
with an accuracy better than 0.05 mag.}
\label{Fig5}
\end{figure}

A glance at the data plotted in Figures 4 and 5, and in particular the 
small color dispersion along the HB and the Red Giant Branch (RGB) show 
that photometry is very accurate down to limiting magnitudes of 
$K\approx19.5$ and $J\approx20.5$. Moreover and even more importantly 
RR Lyrae stars show a well-defined slope both in the J and in the K-band.  

Using the mean K magnitudes provided by ALLFRAME, a mean metallicity 
of [Fe/H]=-1.71 based on spectroscopic data (Suntzeff et al.\cite{ss92}), 
a mean reddening of E(B-V)=0.02 (Walker\cite{w92}), the Cardelli et 
al.\cite{ccm89} relation, and the $PLZ_K$ relation discussed in section 3,  
we found a true distance modulus of $18.45\pm0.04$ mag, where the 
uncertainty only accounts for internal photometric errors. Interestingly 
enough, by adopting the mean J-band magnitude, the same assumptions 
concerning metallicity and reddening, and a new $PLZ_J$ relation, we 
found a distance modulus of $18.51\pm0.06$ mag, where the uncertainty 
only accounts for internal photometric errors.

\section{Anomalous Cepheids}

Anomalous Cepheids are an interesting group of variable stars, since they 
are brighter than RR Lyrae stars and have periods that range from 0.5 
days to a few days. They have been identified booth in GGCs and in LG dwarf 
galaxies (Nemec, Nemec, \& Lutz\cite{nnl94}). Dating back to 
Demarque \& Hirshfeld\cite{dh75} and to Hirshfeld\cite{h80} the common 
belief concerning the evolutionary status 
of these objects is that they are metal-poor, intermediate-mass stars 
with an age of the order of 1 Gyr. This hypothesis was confirmed by 
more recent evolutionary (Castellani \& Degl'Innocenti\cite{cast95}; 
Caputo \& Degl'Innocenti\cite{cd95}) and pulsational  
(Bono et al.\cite{bcs97}) investigations.  
However, the region of the HR diagram roughly located at $\log L/L_\odot=2$ 
presents several intrinsic features worth being discussed in some 
detail. The top panel of Fig. 6 shows the HR diagram for metal-poor,  
intermediate-mass stars ranging from 2.2 to 3.5 $M/M_\odot$. It is 
worth mentioning that the minimum mass that performs the blue loop 
for this composition is $2.2 M/M_\odot$. This mass value is smaller 
than the corresponding minimum mass for the chemical compositions 
typical of the Small 
($3.25 M/M_\odot$, Z=0.004) and of the Large ($4.25 M/M_\odot$, 
Z=0.01) Magellanic Cloud (Bono et al.\cite{bt00}). This means 
that metal-poor stellar systems such as IC1613 should produce a 
substantial fraction of short-period classical Cepheids. This suggestion 
is supported by current empirical evidence (Udalski et al.\cite{uda01}).  
Moreover and even more importantly evolutionary tracks plotted in this 
panel show that the blue loop takes place at hotter effective temperatures 
when moving from 2.2 to 3.5 $M/M_\odot$. The occurrence of this behavior 
was explained by Cassisi \& Castellani\cite{cas93} as the 
consequence of the fact that metal-poor intermediate-mass models do not 
reach the Hayashi track before central helium ignition. Therefore,  
these models do not undergo the canonical dredge-up phase. 
We also note that for evolutionary models more massive than 
$3.5  M/M_\odot$ the amount of time spent inside the instability strip 
is substantially shorter (Pietrinferni et al. 2003, in preparation) when 
compared to more metal-rich models.  

\begin{figure}[ht]
\hspace*{-1.25truecm}\includegraphics[width=1.1\textwidth]{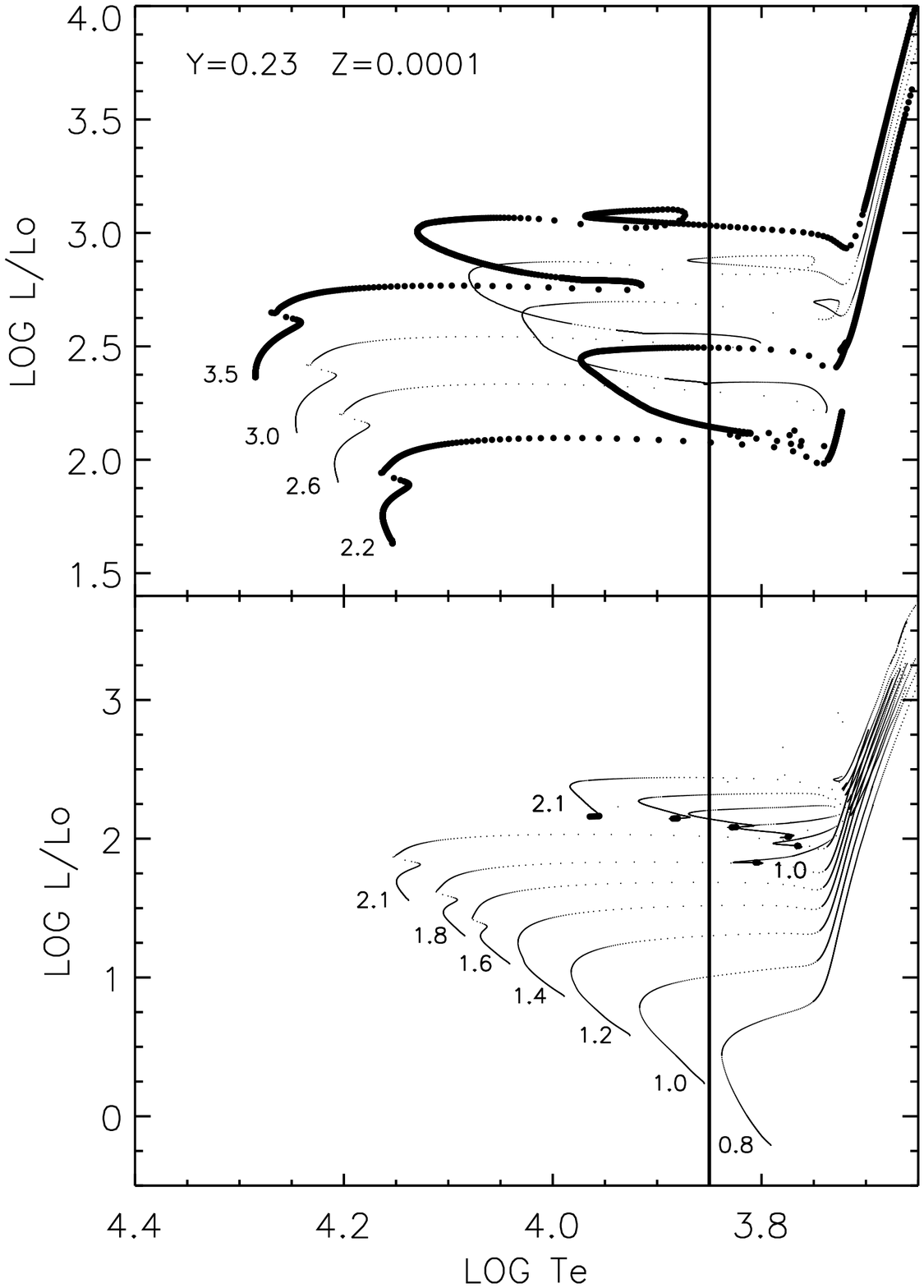}

\vspace*{1.33truecm} 
\caption[]{Top: HR diagram for intermediate-mass stars at fixed chemical 
composition (Y=0.23, Z=0.0001). Note that the blue loop when moving from 
2.2 $M_\odot$ evolutionary models to 3.5 $M_\odot$ takes place at hotter 
effective temperatures. The vertical line marks the center of the 
Cepheid instability strip. The width in temperature of the instability 
strip is typically $\pm0.05$ dex. Bottom: same 
as the top panel but for low and intermediate-mass stars. Filled circles 
mark the beginning of central He-burning phases for models ranging 
from 1.0 $M_\odot$ to 2.1 $M_\odot$. Data plotted in this figure illustrate  
that central He-burning phases take place inside the instability strip 
for evolutionary models more massive than 1.4 $M_\odot$ and less massive 
than 1.0 $M_\odot$.}\label{Fig6}
\end{figure}

Data plotted in the bottom panel of Fig. 6 show quite clearly that 
the structures less massive than 2.2 $M/M_\odot$ show a substantially 
different behavior, and indeed they start to burn helium in the center 
at effective temperatures of the order of $\log T_e =3.9-4.0$. The 
temperature range moves to lower effective temperatures and crosses 
the instability strip for structures with mass values of the order of 
1.4-1.8 $M/M_\odot$. These structures spend a substantial amount of 
He-burning phases inside the instability strip and should produce 
Anomalous Cepheids. The central He-burning phases of less-massive 
structures performs a "hook", i.e. they move at first toward lower effective 
temperatures (1.0-1.2 $M/M_\odot$) and then toward higher effective 
temperatures (0.9-1.0 $M/M_\odot$). Structures with mass values smaller 
than the latter limit produce RR Lyrae stars. This is a very qualitative 
scenario and a more detailed analysis can be found in Castellani \& 
Degl'Innocenti\cite{cast95}. A few {\em caveats} concerning the previous 
observational scnario: {\em i} The evolutionary tracks plotted in 
Fig. 6 have been computed by adopting a Reimers mass-loss rate with 
$\eta=0.4$. This means that mass values that cross the instability 
strip slightly depend on this assumption. {\em ii} Stellar structures 
producing Anomalous Cepheids and classical Cepheids show a substantial 
difference in the age range covered by main sequence stars. In fact, 
evolutionary models with stellar mass $\approx 1.6 M/M_\odot$ spend 
on the main sequence a lifetime of roughly 0.7 Gyr, while models of 
3.0 $M/M_\odot$ leave the main sequence after approximately 0.2 Gyr. 
This means that stellar systems producing classical Cepheids should 
also show a well-populated blue main sequence region when compared 
with stellar systems producing Anomalous Cepheids.  
{\em iii} Current evolutionary scenario for Anomalous Cepheids relies 
on the assumption that they are the aftermath of single star evolution. 
However, the occurrence of a few Anomalous Cepheids in GGCs suggest 
that a fraction of them might be the progeny of binary collisions or 
of binary mergings (Renzini\cite{ar80}; Nemec et al.\cite{nnl94}; 
Bono et al.\cite{bcs97}).

\begin{figure}[ht]
\hspace*{-1.25truecm}\includegraphics[width=1.1\textwidth]{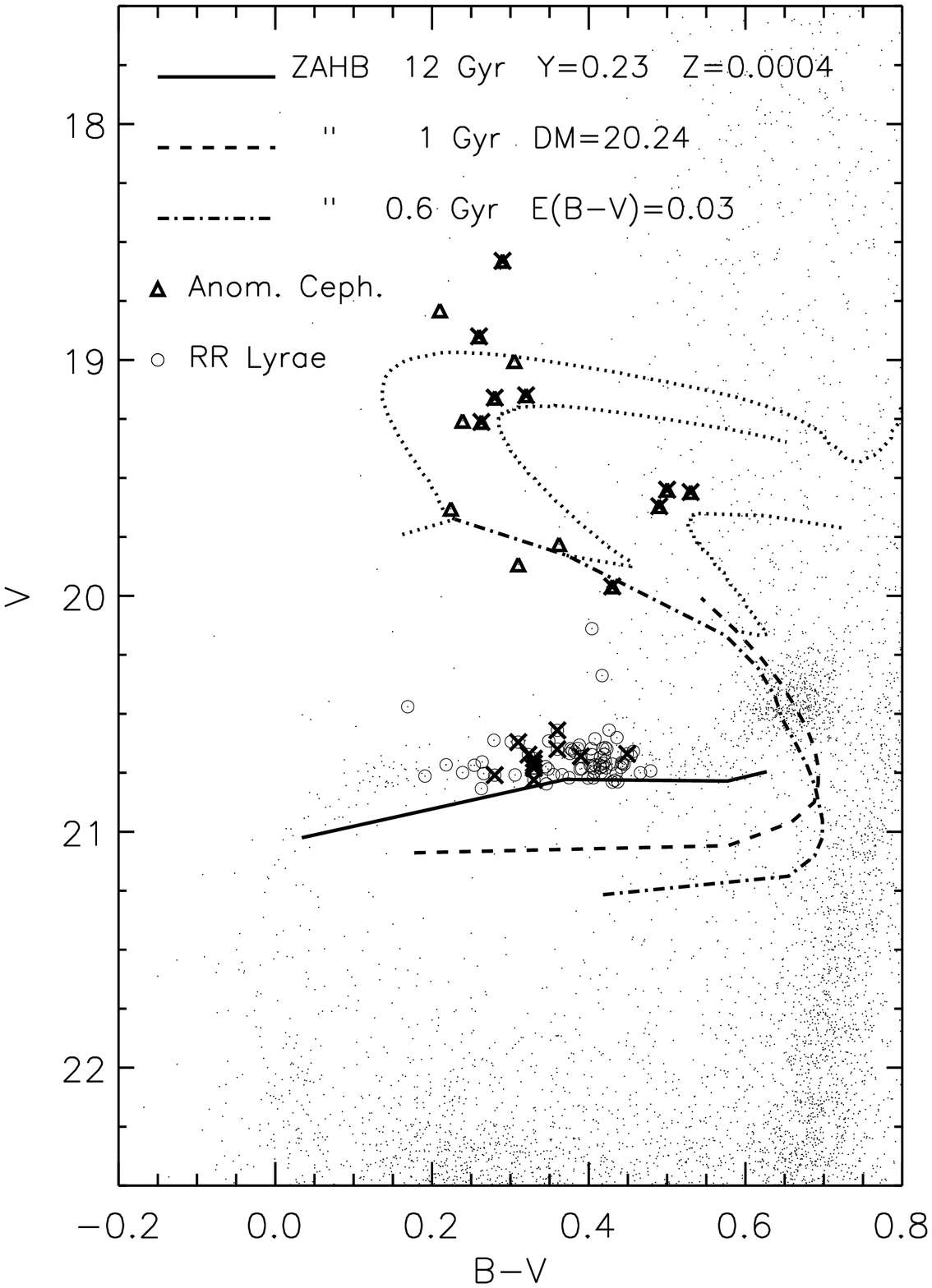}

\vspace*{1.33truecm} 
\caption[]{Comparison between predicted He-burning structures at 
fixed chemical composition (Z=0.0004, Y=0.23) and bright stars in 
the Carina dwarf galaxy (Dall'Ora et al.\cite{mda03}; 
Monelli et al.\cite{mmo03}). 
Data plotted in this figure show static (small dots) and variable 
stars:  circles RR Lyrae stars, triangles, ACs. Crosses mark
variables that present poor-phase coverage. Solid, dashed, and dotted-dashed 
lines display predicted Zero Age He-burning structures for different 
progenitor ages ranging from 12 ($M=0.8 M_\odot$) to 0.6 ($M=2.2 M_\odot$) 
Gyr. The dotted lines show the He-burning evolution for three 
intermediate-age structures of 1.8 (redder), 2.0, and 2.2 (bluer) 
$M/M_\odot$.}\label{Fig7}
\end{figure}

The observational scenario concerning Anomalous Cepheids has been 
substantially improved during the last few years (Siegel \& 
Majewski\cite{sie00}; Bersier \& Wood\cite{ber02}; 
Dolphin et al.\cite{d02};  Pritzl et al.\cite{pr02}). 
Fig. 7 shows the distribution of both RR Lyrae and Anomalous Cepheids 
detected by Dall'Ora et al.\cite{mda03} in the Carina dwarf galaxy. 
The comparison between theory and observations supports the evolutionary 
scenario we discussed in this section. In fact, Dall'Ora et al.\cite{mda03} 
and Monelli et al.\cite{mmo03} found, on the basis of pulsational and 
evolutionary 
arguments that these objects are approximately a factor of two more 
massive than RR Lyrae stars present in the same galaxy. Note that the 
metal abundance adopted for this stellar system is Z=0.0004. 
Interestingly enough,  evolutionary predictions also suggest that 
more metal-rich structures should not produce Anomalous Cepheids, since 
the so-called "hook" of intermediate-mass helium burning structures do 
not cross the instability strip.  

Theory and observations suggest that Anomalous Cepheids pulsate both in 
the fundamental and in the first overtone (Nemec, et al.\cite{nnl94}; 
Bono et al.\cite{bcs97}). However, more data are required to 
constrain on a 
quantitative basis the accuracy of distance determinations based on 
these objects (Pritzl et al.\cite{pr02}). Obviously LG dwarf 
galaxies are crucial systems to investigate this problem, since 
several of them host both RR Lyrae, Anomalous Cepheids, and large 
samples of red clump stars.

\section{Conclusions}

The results concerning the RR Lyrae distance scale discussed in this 
investigation can be summarized along two different paths: 

{\em Theoretical path.} a) Theoretical predictions based on pulsation 
models, namely the $PLZ_K$ relation and the FOBE method supply, within 
current uncertainties, similar absolute distances. The distance 
to M3 provided by the former method is in very good agreement with the 
empirical calibration provided by Longmore et al.\cite{l90}. Moreover, 
the pulsation parallax obtained for RR Lyr itself is in remarkable 
agreement with the trigonometric parallax recently obtained by Benedict 
et al.\cite{ben02}. These findings, once confirmed by independent 
investigations, 
together with plain physical arguments concerning the dependence of
the bolometric correction and of the color-temperature relation on 
input physics indicate that the $PLZ_K$ might be less affected by 
deceptive uncertainties affecting the other methods. This approach 
can be further improved. Up to now theoretical and empirical $PLZ_K$ 
relations were derived by simultaneously accounting for $RRab$ and $RRc$  
variables. First overtones (FOs) are "fundamentalized" by adding 0.13 to 
the logarithm of the period. However, preliminary theoretical results 
suggest that FOs also obey a well-defined $PLZ_K$ relation. The main 
advantage in using FOs is that the instability region of these 
objects is narrower when compared with fundamental mode RR Lyrae. 
Therefore the FO $PLZ_K$ relation presents a smaller intrinsic dispersion.  

b) Theoretical predictions based on evolutionary models have been 
widely discussed in the literature (Bono, Castellani, \& Marconi\cite{bcm00};
Cassisi et al.\cite{cas99}; Caputo et al.\cite{ccm00}). The main outcome 
of these 
investigations is that current HB models seem to predict HB luminosities 
that are $\approx 0.1$ mag brighter than estimated using the pulsational 
approach. However, different sets of HB models constructed by adopting 
different assumptions on input physics present a spread in HB luminosities 
of the order of 0.15 mag. This means that in the near future new 
observational constraints based either on geometrical distances or on 
robust distance indicators might supply the unique opportunity to 
nail down the intrinsic accuracy of the ingredients currently adopted 
in evolutionary and pulsation models.  

c) In section 2, we mentioned that we still lack homogeneous sets 
of atmosphere models that cover a broad range of microturbulent 
velocities. New models are strongly required to check on a quantitative 
basis the impact that such a parameter has on the transformation 
of theoretical predictions into the observational plane. The new 
models might also play a crucial role in understanding the plausibility 
of the physical assumptions currently adopted by the BW method.

{\em Observational path.} a) Theoretical and empirical evidence suggests  
that the $PLZ_K$ relation for RR Lyrae presents several advantages when 
compared with other methods available in the literature. This 
notwithstanding we still lack of accurate K-band measurements for both 
field and cluster RR Lyrae. The use of current generation NIR detectors 
at 4m class telescopes and careful reduction strategies seem to suggest 
that accurate mean K-band magnitudes can be obtained down to 
$K\approx18.5-19.0$. This means that we should be able to supply 
an accurate distance scale for Galactic and Large Magellanic cloud 
clusters. In the near future the use of NIR detectors at 8m class 
telescopes should allow us to detect and measure RR Lyrae stars in 
several Local Group (LG) galaxies (see the ARAUCARIA project,
Pietrinzski et al. in these proceedings). This is a fundamental step 
to improve the global accuracy of cosmic distances, because LG galaxies 
display complex star formation histories, and often host not 
only RR Lyrae but also intermediate-mass distance indicators, such 
as Red Clump stars, Anomalous Cepheids, and classical Cepheids.  
 
b) We focused our attention on the mean K-band magnitude of RR Lyrae 
stars. However, theory and observations suggest that RR Lyrae do 
obey a PL relation also in the J and the H-band. Once again the 
amount of data available in the literature for these bands is quite
limited.

\section{Acknowledgements}
It is a pleasure to the thank several collaborators with whom I share 
this intriguing ongoing effort aimed at understanding evolutionary 
and pulsational properties of radial variables across the HR diagram.    
Special thanks to S. Cassisi and S. Degl'Innocenti for sending me 
evolutionary tracks for low and intermediate-mass stars in advance 
of publication.  
I am very grateful to V. Castellani, F. Caputo, H. Smith, and A. Walker
for a detailed reading of an early draft of this paper and for many 
enlightning discussions and suggestions.  
This work was partially supported by MIUR/Cofin~2001 under the project: 
"Origin and Evolution of Stellar Populations in the Galactic Spheroid"
and by MIUR/Cofin~2002 under the project: "Stellar Population in Local 
Group Galaxies".


%

\end{document}